\documentclass[sigconf,natbib=false,authorversion]{acmart}

\AtBeginDocument{\providecommand\BibTeX{{\normalfont B\kern-0.5em{\scshape i\kern-0.25em b}\kern-0.8em\TeX}}}

\copyrightyear{2023}
\acmYear{2023}
\setcopyright{rightsretained}
\acmConference[CF '23]{20th ACM International Conference on Computing
Frontiers}{May 9--11, 2023}{Bologna, Italy}
\acmBooktitle{20th ACM International Conference on Computing Frontiers (CF
'23), May 9--11, 2023, Bologna, Italy}
\acmDOI{10.1145/3587135.3592175}
\acmISBN{979-8-4007-0140-5/23/05}

\usepackage{float}
\usepackage[caption=false]{subfig}
\usepackage[subtle]{savetrees}

\usepackage{amsmath,amsfonts}
\usepackage{algorithmic}
\usepackage{graphicx}
\usepackage{textcomp}
\usepackage{xcolor}
\usepackage{paralist}
\usepackage{xspace}
\usepackage{hyperref}
\usepackage[inkscapeformat=png]{svg}
\usepackage{orcidlink}
\usepackage[backend=biber,style=acmnumeric,minnames=1,maxnames=1,giveninits=true]{biblatex}

\usepackage{xpatch}

\makeatletter
\patchcmd\blx@bblinput{\blx@blxinit}
                      {\blx@blxinit
                      }{}{\fail}
\makeatother
			     
\AtBeginBibliography{\footnotesize}
\AtEveryBibitem{\clearfield{pages}\clearlist{location}\clearlist{organization}\clearfield{pagetotal}\clearfield{eprint}\clearfield{doi}\clearfield{isbn}\clearfield{number}\clearfield{volume}\clearfield{note}\clearfield{issn}\clearfield{series}\clearfield{editor}\clearlist{editor}\ifentrytype{online}{}{\clearfield{url}}}
\renewbibmacro*{year}{\iffieldundef{year}{}{\printfield{year}}}
\usepackage{pifont}
\usepackage{siunitx}
\usepackage{placeins}
\usepackage{csquotes}
\usepackage{listings}

\definecolor{mGreen}{rgb}{0,0.6,0}
\definecolor{mGray}{rgb}{0.5,0.5,0.5}
\definecolor{mPurple}{rgb}{0.58,0,0.82}
\definecolor{backgroundColour}{rgb}{0.95,0.95,0.92}

\def\BibTeX{{\rm B\kern-.05em{\sc i\kern-.025em b}\kern-.08em
    T\kern-.1667em\lower.7ex\hbox{E}\kern-.125emX}}

\makeatletter
\begin{document}

\title[VEDLIoT --- Next generation accelerated AIoT systems and applications]{VEDLIoT --- Next generation\texorpdfstring{\\}{} accelerated AIoT systems and applications}
\subtitle{Invited Paper}

\author{\href{mailto:kmika@techfak.uni-bielefeld.de}{Kevin Mika}~\orcidlink{0009-0005-2717-5147}}
\author{\href{mailto:rgriessl@techfak.uni-bielefeld.de}{Ren\'e Griessl}~\orcidlink{0009-0004-5565-9141}}
\author{\href{mailto:nkucza@techfak.uni-bielefeld.de}{Nils Kucza}~\orcidlink{0009-0000-7963-0280}}
\author{\href{mailto:fporrmann@techfak.uni-bielefeld.de}{Florian Porrmann}~\orcidlink{0000-0003-2401-7862}}
\affiliation{\institution{\href{https://www.cit-ec.de/en/ks}{Bielefeld University}}
  \city{Bielefeld}
  \country{Germany}
}
\author{\href{mailto:mkaiser@techfak.uni-bielefeld.de}{Martin Kaiser}~\orcidlink{0009-0001-9854-027X}}
\author{\href{mailto:ltigges@techfak.uni-bielefeld.de}{Lennart Tigges}~\orcidlink{0009-0000-7339-2541}}
\author{\href{mailto:jhagemey@techfak.uni-bielefeld.de}{Jens Hagemeyer{\normalsize\textsuperscript{*}}~\orcidlink{0009-0005-9943-8081}}
\authornote{Corresponding author: Jens Hagemeyer, e-mail: \href{mailto:jhagemey@techfak.uni-bielefeld.de}{jhagemey@techfak.uni-bielefeld.de}
}}
\affiliation{\institution{\href{https://www.cit-ec.de/en/ks}{Bielefeld University}}
  \city{Bielefeld}
  \country{Germany}
}

\author{\href{mailto:ppedro@chalmers.se}{Pedro~Trancoso}~\orcidlink{0000-0002-2776-9253}}
\author{\href{mailto:waqarm@chalmers.se}{Muhammad Waqar Azhar}~\orcidlink{0000-0003-0477-4540}}
\author{\href{mailto:qarayah@chalmers.se}{Fareed Qararyah}~\orcidlink{0000-0002-3955-2836}}
\author{\href{mailto:zouzoula@chalmers.se}{Stavroula Zouzoula}~\orcidlink{0009-0009-4419-6491}}
\affiliation{\institution{\href{https://www.chalmers.se}{Chalmers University of Technology}}
  \city{Gothenburg}
  \country{Sweden}
}

\author{\href{mailto:james.menetrey@unine.ch}{J\"ames M\'en\'etrey}~\orcidlink{0000-0003-2470-2827}}
\author{\href{mailto:marcelo.pasin@unine.ch}{Marcelo Pasin}~\orcidlink{0000-0002-3064-5315}}
\author{\href{mailto:pascal.felber@unine.ch}{Pascal Felber}~\orcidlink{0000-0003-1574-6721}}
\affiliation{\institution{\href{http://www.unine.ch/iiun/home/chaires-de-recherche/systemes-complexes.html}{University of Neuch\^{a}tel}}
  \city{Neuch\^{a}tel}
  \country{Switzerland}
}

\author{\href{mailto:carina.marcus@veoneer.com}{Carina Marcus}~\orcidlink{0000-0003-4147-1805}}
\author{\href{mailto:oliver.brunnegard@veoneer.com}{Oliver Brunnegard}~\orcidlink{0000-0003-1266-4173}}
\author{\href{mailto:olof.eriksson@veoneer.com}{Olof Eriksson}~\orcidlink{0009-0009-4182-9222}}
\affiliation{\institution{\href{https://www.veoneer.com}{VEONEER Inc.}}
  \city{V\r{a}rg\r{a}rda}
  \country{Sweden}
}

\author{\href{mailto:hans@embedl.ai}{Hans Salomonsson}~\orcidlink{0000-0002-9615-7410}}
\author{\href{mailto:daniel@embedl.ai}{Daniel \"Odman}~\orcidlink{0009-0009-3979-1149}}
\author{\href{mailto:andreas@embedl.com}{Andreas Ask}~\orcidlink{0009-0001-7003-2409}}
\affiliation{\institution{\href{https://embedl.ai}{EMBEDL AB}}
  \city{Gothenburg}
  \country{Sweden}
}

\author{\href{mailto:casim@ciencias.ulisboa.pt}{Antonio Casimiro}~\orcidlink{0000-0002-5522-5739}}
\author{\href{mailto:anbessani@fc.ul.pt}{Alysson Bessani}~\orcidlink{0000-0002-8386-1628}}
\author{\href{mailto:tcarvalho@lasige.di.fc.ul.pt}{Tiago Carvalho}~\orcidlink{0000-0002-2242-8265}}
\affiliation{\institution{\href{http://www.di.fc.ul.pt/}{University of Lisbon}}
  \city{Lisbon}
  \country{Portugal}
}

\author{\href{mailto:kgugala@antmicro.com}{Karol Gugala}~\orcidlink{0009-0007-6927-6370}}
\author{\href{mailto:pzierhoffer@antmicro.com}{Piotr Zierhoffer}~\orcidlink{0009-0001-5852-7644}}
\author{\href{mailto:glatosinski@antmicro.com}{Grzegorz Latosinski}~\orcidlink{0009-0001-4336-3499}}
\affiliation{\institution{\href{https://antmicro.com}{Antmicro}}
  \city{Pozna\'n}
  \country{Poland}
}

\author{\href{mailto:marco.tassemeier@uni-osnabrueck.de}{Marco Tassemeier}~\orcidlink{0000-0001-9498-8416}}
\author{\href{mailto:mporrmann@uni-osnabrueck.de}{Mario Porrmann}~\orcidlink{0000-0003-1005-5753}}
\affiliation{\institution{\href{https://www.inf.uni-osnabrueck.de/research_groups/computer_engineering.html}{Osnabr\"uck University}}
  \city{Osnabr\"uck}
  \country{Germany}
}

\author{\href{mailto:hans-martin.heyn@gu.se}{Hans-Martin Heyn}~\orcidlink{0000-0002-2427-6875}}
\author{\href{mailto:eric.knauss@cse.gu.se}{Eric Knauss}~\orcidlink{0000-0002-6631-872X}}
\affiliation{\institution{\href{https://www.gu.se/en}{University of Gothenburg}}
  \city{Gothenburg}
  \country{Sweden}
}

\author{\href{mailto:yufei.mao@siemens.com}{Yufei Mao}~\orcidlink{0009-0003-3211-7748}}
\author{\href{mailto:franz.meierhoefer@siemens.com}{Franz Meierh\"ofer}~\orcidlink{0009-0002-5873-611X}}
\affiliation{\institution{\href{https://www.siemens.com}{Siemens AG}}
  \city{Erlangen}
  \country{Germany}
}
\renewcommand{\shortauthors}{Mika et al.}

\begin{abstract}
The VEDLIoT project aims to develop energy-efficient Deep Learning methodologies for distributed Artificial Intelligence of Things (AIoT) applications. During our project, we propose a holistic approach that focuses on optimizing algorithms while addressing safety and security challenges inherent to AIoT systems. The foundation of this approach lies in a modular and scalable cognitive IoT hardware platform, which leverages microserver technology to enable users to configure the hardware to meet the requirements of a diverse array of applications. Heterogeneous computing is used to boost performance and energy efficiency. In addition, the full spectrum of hardware accelerators is integrated, providing specialized ASICs as well as FPGAs for reconfigurable computing. The project's contributions span across trusted computing, remote attestation, and secure execution environments, with the ultimate goal of facilitating the design and deployment of robust and efficient AIoT systems. The overall architecture is validated on use-cases ranging from Smart Home to Automotive and Industrial IoT appliances. Ten additional use cases are integrated via an open call, broadening the range of application areas.
\end{abstract}

\ccsdesc[300]{Hardware~Hardware accelerators}
\ccsdesc[300]{Hardware~Reconfigurable logic applications}
\ccsdesc[300]{Computer systems organization~Embedded hardware}
\ccsdesc[100]{Computer systems organization~Reliability}
\ccsdesc[300]{Computer systems organization~Neural networks}
\ccsdesc[300]{Computer systems organization~Reconfigurable computing}
\ccsdesc[300]{Computer systems organization~Heterogeneous (hybrid) systems}

\keywords{Artificial Intelligence of Things (AIoT), Acceleration, Reconfigurable and Heterogeneous Computing, Machine Learning (ML), Distributed Attestation and Security}

\renewcommand{\shortauthors}{Mika et al.}
\def\authors{\shortauthors}
\maketitle
\renewcommand{\shortauthors}{Mika et al.}

\section{Introduction}
\label{sec:introduction}
Deep Learning (DL) has become a strong driver in IoT applications. Such applications usually have challenging computational and memory requirements, coupled with a low energy budget. VEDLIoT aims at enabling the use of DL algorithms in IoT by accelerating and optimizing applications with energy efficiency in mind. An overview of the project was given in a previous publication~\cite{VEDLIoT-date22}, providing a basis for this one. 
\begin{figure}[h]
\centering
\includegraphics[width=0.93\columnwidth]{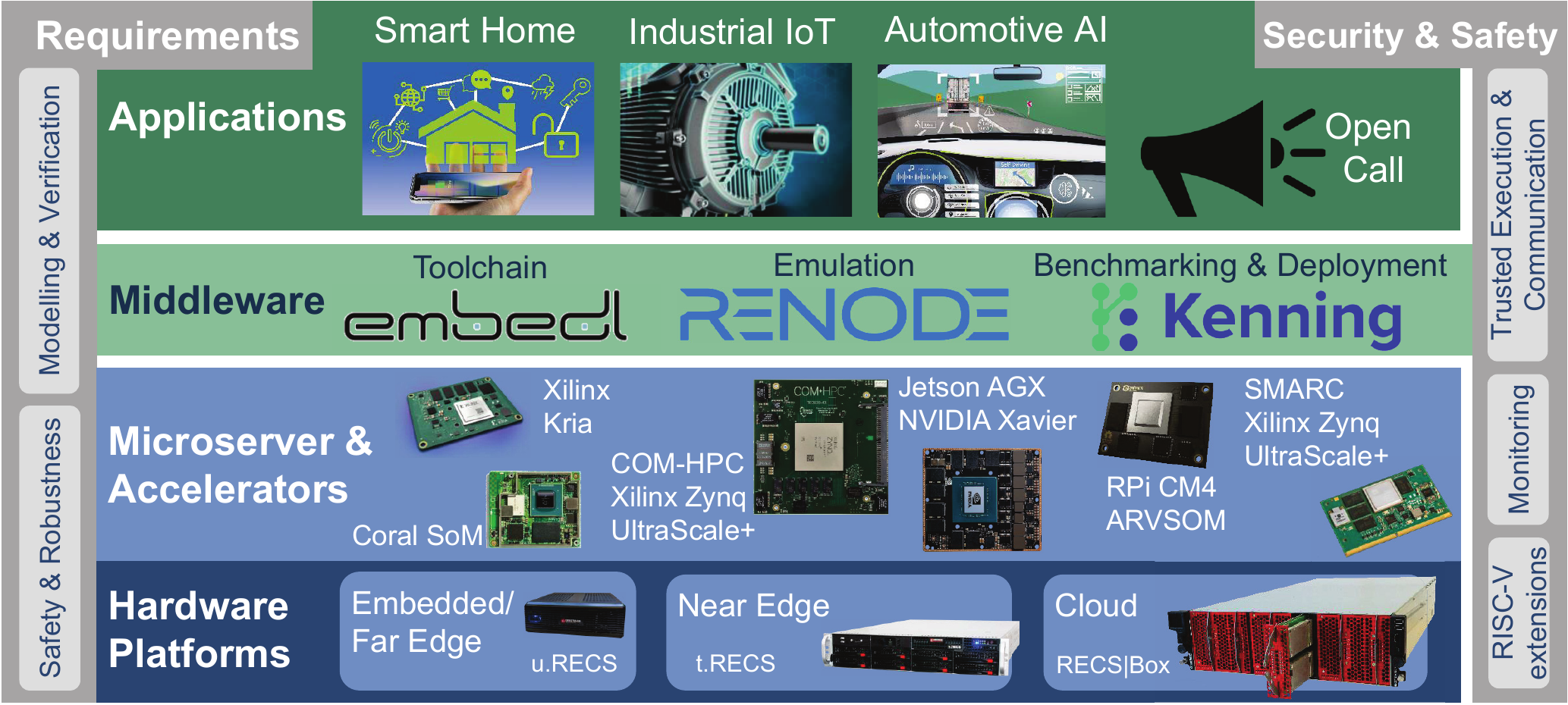} 
\caption{VEDLIoT architecture overview}
\label{fig:vedliot_overview}
\vspace{-1.5em}
\end{figure}
Here, we give more insights regarding the outcomes and advances of VEDLIoT aspects from bottom to top (\autoref{fig:vedliot_overview}), starting with the newly developed u.RECS microserver hardware platform to the toolchains and use cases. Security aspects and requirements engineering accompany throughout the development.

 \section{Accelerated AIoT Hardware\texorpdfstring{\\}{ }Platform}
\label{sec:hwplatforms}
In this section, the VEDLIoT cognitive AIoT hardware platform as well as benchmarking results regarding YoloV4 of the different microservers are presented in a compressed manner. A more in-depth presentation of the architecture, with performance measurements for many different accelerators (CPU, GPU, ASCI, FPGA) can be found in a previous publication~\cite{VEDLIoT-date23}.
\subsection{Heterogeneous hardware platform}
RECS represents a heterogeneous hardware platform that has been used and refined in various EU-projects. RECS is a flexible microserver architecture that can accommodate a range of computing elements, including x86, 64-bit ARM mobile/embedded processors, 64-bit ARM server processors, FPGAs, GPUs, and ASICs. Due to this nature it can be upgraded and altered to fit the use case specific requirements~\cite{griessl2014scalable,legato-date}. Unlike traditional microserver platforms which support only homogeneous devices, RECS allows for the seamless integration of diverse technologies, enabling fine-tuning of the platform towards specific applications, providing a comprehensive cloud-to-edge platform.

\begin{figure}[t]
\centering
\includegraphics[width=0.75\columnwidth]{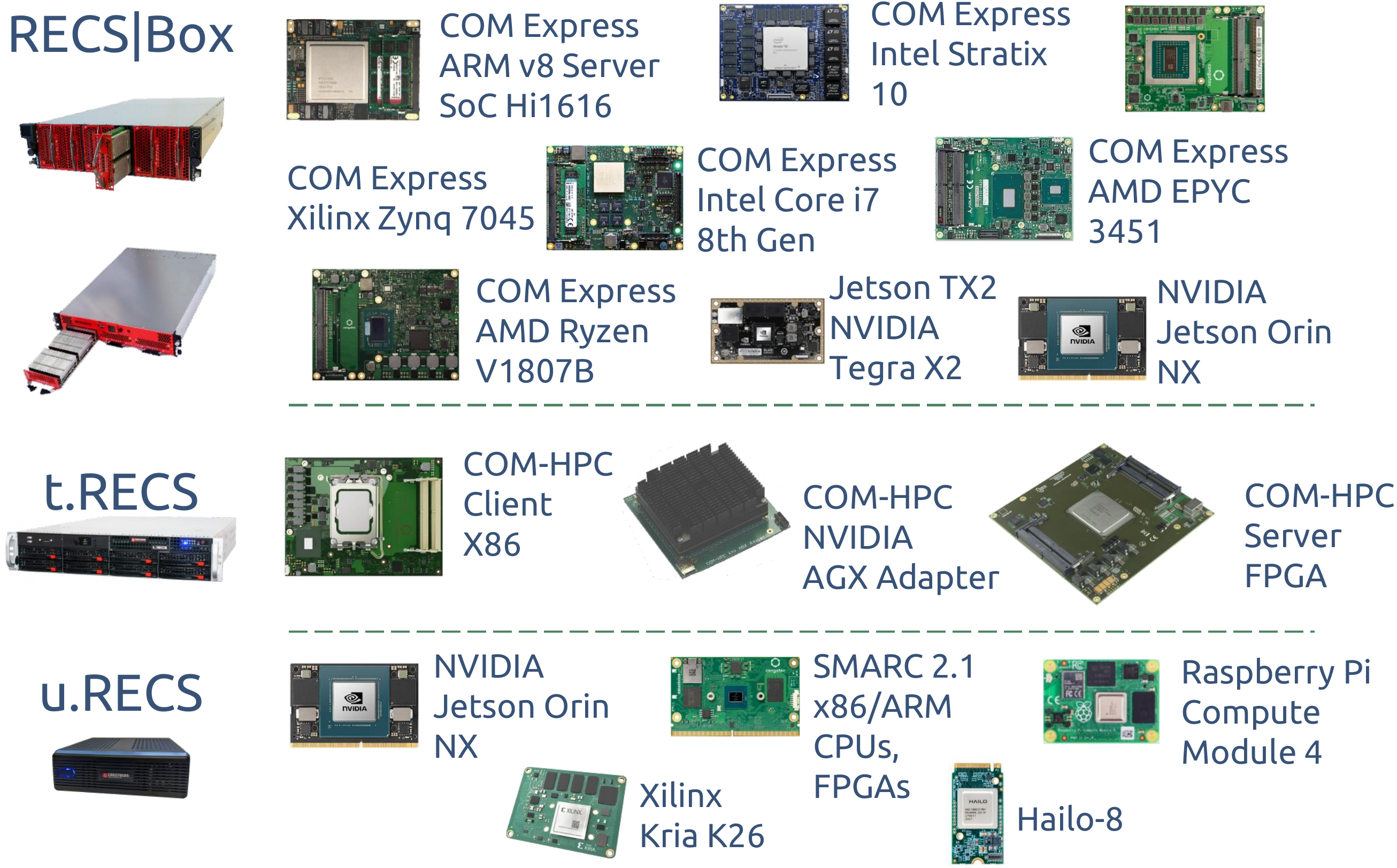} 
\caption{Overview of RECS and the different microservers}
\label{fig:recs_overview}
\vspace{-2.0em}
\end{figure}

This results in a densely-coupled, highly-integrated heterogeneous microserver, with a high-speed, low-latency communication infrastructure. \autoref{fig:recs_overview} gives an overview over the RECS system, with the RECS{\textbar}Box on top and the u.RECS at the bottom, also a selection of fitting microservers is pictured.

The RECS family includes several systems, supporting efficient computing from cloud to edge, with the most powerful being the RECS{\textbar}Box, an HPC and cloud system, capable of hosting up to 27 COM Express microservers, 144 Nvidia Jetson microservers, or a custom mix of these components. It is a 19” microserver system for heterogeneous CPU, GPU and FPGA computing, featuring PCIe with high-speed, low-latency communication infrastructure~\cite{oleksiak2017m2dc}.

For edge applications such as 5G base stations or similar the t.RECS is available. It boasts various advanced features, including three COM-HPC module slots, switched 10~Gbps Ethernet, and integrated i-KVM for remote management. Additionally, the t.RECS features an integrated PCIe switch that enables fast, low-latency communication between host and target devices. One of the key design features of the t.RECS is its ability to accommodate up to three COM-HPC microservers, divided into two client slots and one server slot. 

The u.RECS, the smallest RECS family member, is designed for far-edge scenarios, such as autonomous vehicles or production facilities, facilitating local AI workload computation. It comprises three module slots: an NVIDIA NX slot for embedded GPUs (e.g., NVIDIA Orin NX), a SMARC 2.1 slot for diverse modules (e.g., FPGAs, x86 or ARM processors), and an M.2 slot for dedicated AI accelerators (e.g., Hailo-8 or Intel Myriad). Adapters enable compatibility with Raspberry Pi CM4 or Xilinx Kria, while an mPCIe slot supports 5G, Wi-Fi, or similar extensions. Configurable PCIe connections allow for adaptable topologies, with 1~Gbps Ethernet connectivity between modules, exposed via two PoE-enabled RJ45 ports, that can be used, e.g., to connect Ethernet cameras. A 1~Gbps single-pair Ethernet connection is available for automation and automotive use cases. \subsection{Accelerators and microservers}
\label{sec:hwplatforms:acc}
In VEDLIoT, performance and energy efficiency assessments are critical in selecting DL accelerators for RECS integration, ensuring customization for distinct use cases. The project examines various architectures, including GPUs, FPGAs, and ASICs, as a burgeoning market offers diverse DL hardware accelerators for various applications. This evaluation helps hardware selection within and beyond the project.

\begin{figure}[b]
\centering
\includegraphics[width=0.99\columnwidth]{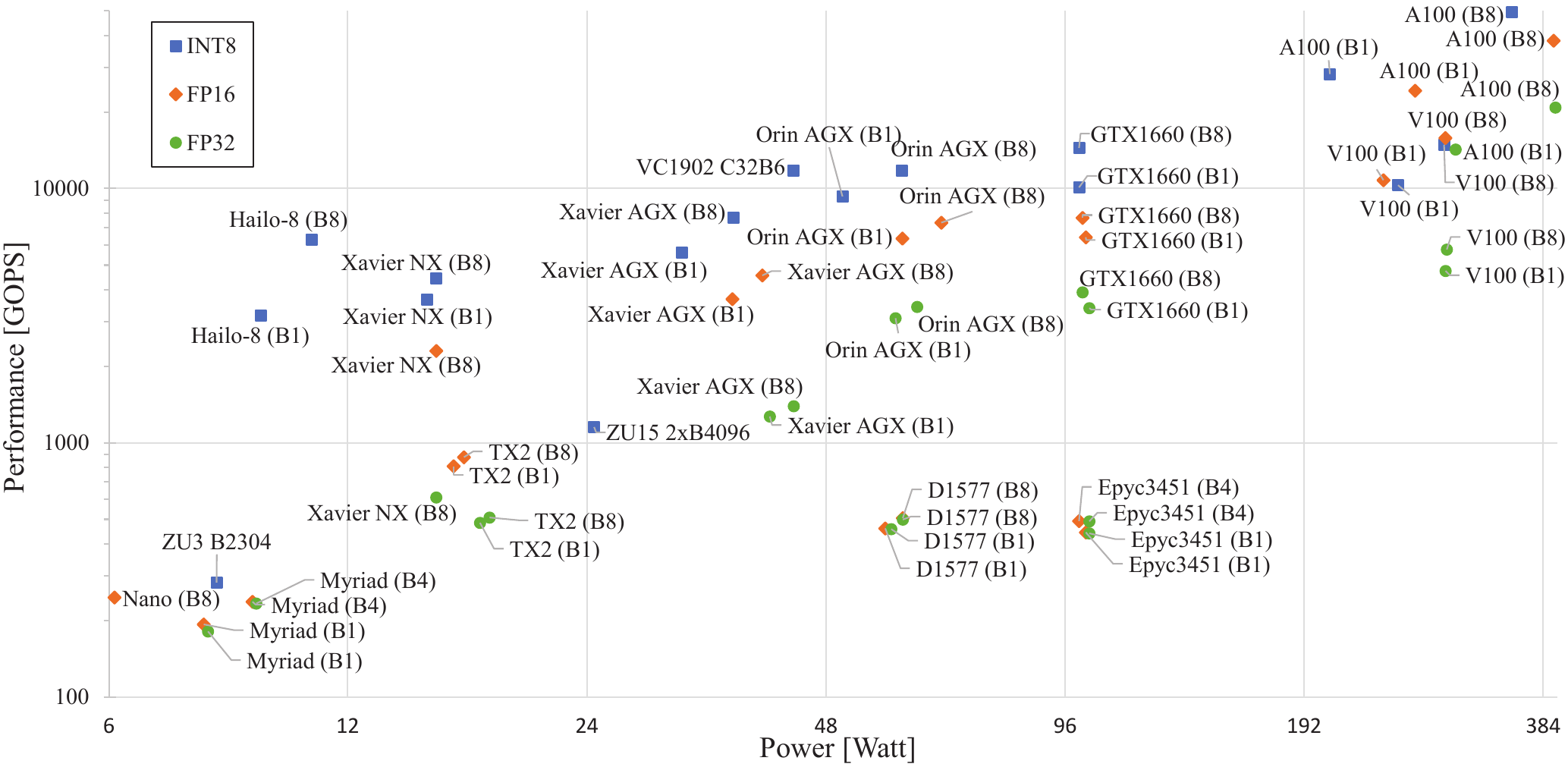} 
\caption{YoloV4 performance evaluation of DL accelerators}
\label{fig:hardware_yolo}
\vspace{-2.0em}
\end{figure}

To assess performance, YoloV4 models were employed to compare available accelerators. Tests used INT8, FP16, or FP32 datatypes based on hardware quantization support, with optimal tools chosen per manufacturer recommendations (e.g., TensorRT for NVIDIA). Batch sizes from 1 to 8 (B1, B4, B8) optimized performance and utilization. \autoref{fig:hardware_yolo} displays YoloV4 performance (in GOPS) and power consumption (in Watt) across platforms, including x86 CPUs (Epic3451, Xeon D1577), GPUs (A100, V100, GTX1660), eGPUs (Xavier AGX, Orin AGX, Xavier NX, Jetson TX2, Jetson Nano), FPGAs/ACAPs (Zynq ZU15, ZU3, Versal VC1902), and ASICs (Hailo-8, Myriad). The results show the full spectrum of different processing architectures from CPUs to modern ASIC accelerators like the Hailo-8 (here measured including Xavier~NX as host system), providing an average efficiency of around 100~$\frac{\textnormal{\scriptsize{GOPS}}}{\textnormal{\scriptsize{Watt}}}$ with up to 1250~$\frac{\textnormal{\scriptsize{GOPS}}}{\textnormal{\scriptsize{Watt}}}$ for Hailo-8.
 \section{Optimizing Toolchain for\texorpdfstring{\\}{ }Heterogeneous Hardware}
\label{sec:toolchain}
\subsection{Optimization of DL models}
\label{sec:toolchain:model}
Efficiency is paramount when deploying DL solutions in IoT contexts. DL systems must fit within memory constraints while maintaining performance in accuracy metrics. VEDLIoT addresses this through hardware-aware pruning and quantization, accelerating DL models and reducing memory footprint without significant accuracy loss. Model compression also enhances energy efficiency, crucial for battery-powered IoT devices.

Hardware speedup is achieved via structured pruning~\cite{molchanov2017pruning}, as opposed to unstructured (sparse) pruning, since embedded hardware often lacks support for acceleration via sparsity. Pruning methods vary based on targeted hardware, as efficient operations or architectures may differ across devices. Embedl's Model Optimization SDK~\cite{embedl} is utilized for hardware-aware pruning.

\subsection{Hardware software co-design}
\label{sec:toolchain:codesign}
The co-design process examines architectural techniques for efficient DL algorithm deployment~\cite{hao2018deep}, utilizing reconfigurable FPGAs to accommodate evolving CNN models. Three primary directions are explored: \emph{(1)} developing an FPGA base design to support the RECS platform, facilitating integration of FPGA-based accelerators~\cite{VEDLIoT-date22}; \emph{(2)} enabling partial dynamic reconfiguration for scenarios necessitating DL accelerator switching or power budgets changes; and \emph{(3)} creating custom and model-specific accelerator designs to enhance targeted CNN efficiency~\cite{qararyah2022fibha}.

Resource-efficient CNNs, also known as heterogeneous, compact, or edge CNNs~\cite{boroumand2021google, xu2021hesa}, minimize computational and memory requirements while maintaining accuracy. VEDLIoT also targets compact CNNs such as MobileNetV3~\cite{howard2019searching}, as monolithic accelerators prove inefficient for these models~\cite{boroumand2021google, xu2021hesa}. Hence, we developed FiBHA (Fixed Budget Hybrid CNN accelerator) that improves over the existing custom accelerators by capturing the heterogeneity of these compact CNNs using a reasonable resource budget~\cite{qararyah2022fibha}. FiBHA employes dedicated engines for heterogeneity-rich layers and minimal engines for less heterogeneous layers. FiBHA outperforms state-of-the-art FPGA-based CNN accelerators like FINN~\cite{blott2018finn}, achieving up to 1.7x and 4.1x throughput improvements. Additionally, VEDLIoT developed FPGA-based accelerators for reinforcement learning~\cite{Rothmann2022} and the STANN framework for designing FPGA-based DNN accelerators.

\subsection{Model verification}
\label{sec:toolchain:ver}

After training, deep neural network models can be subjected to various optimizations for runtime purposes. Firstly, various model compression techniques can be employed to reduce models’ resource and computational demands through e.g. (1) quantization (reduction of weights’ and activations’ precision to 16-bit or 8-bit values), (2) pruning (removing insignificant weights, filters and activations), (3) encoding weights to more compact form (e.g. via clustering). By combining the above techniques it is possible to significantly reduce the model size and inference time, with negligible decrease in quality \cite{han2015deep}. The above compression techniques are often paired with low-level optimizations specific to given hardware platforms, including usage of dedicated accelerators.

Having so many levels of possible optimizations (both on software and hardware level), it is crucial to have a reproducible and empirical flow verifying the correctness of processing, as well as the overall quality of the deployed model. The correctness verification can be easily achieved by using the same evaluation methods that are used during the training process.

Kenning~\cite{kenning} is an open-source framework offering tools for integrating existing frameworks to load, optimize, deploy, and evaluate deep neural networks on target hardware in a traceable and reproducible manner. Once models are optimized and deployed, Kenning automatically evaluates them using test datasets, collecting performance (inference time, memory usage) and quality (accuracy, precision) metrics. Summary reports can be generated for individual or multiple models, including textual and visual comparisons.

\subsection{Simulation}
\label{sec:toolchain:sim}

An FPGA-based ML accelerator using the CFU (Custom Function Unit) RISC-V extension~\cite{cfu-spec} was developed, employing the open-source Renode simulation framework~\cite{renode} for implementation, testing, and debugging. Renode simulates complex, multi-node scenarios, running unmodified software on various architectures (e.g., Arm, RISC-V). To co-simulate Verilog-implemented CFU accelerators, Renode was enhanced to support the CFU interface. CFU code is compiled using Verilator~\cite{verilator} and connected to Renode for mixed functional and cycle-accurate simulation. Additional improvements to bus implementations were made and published alongside usage examples~\cite{renode-verilator}. \section{Safety, Security and Requirements for Distributed AIoT Systems}
\label{sec:safe_sec}

\subsection{Requirements concepts for AIoT}
\label{sec:safe_sec:req}
\begin{figure}[!h]
\centering
\includegraphics[width=0.83\columnwidth]{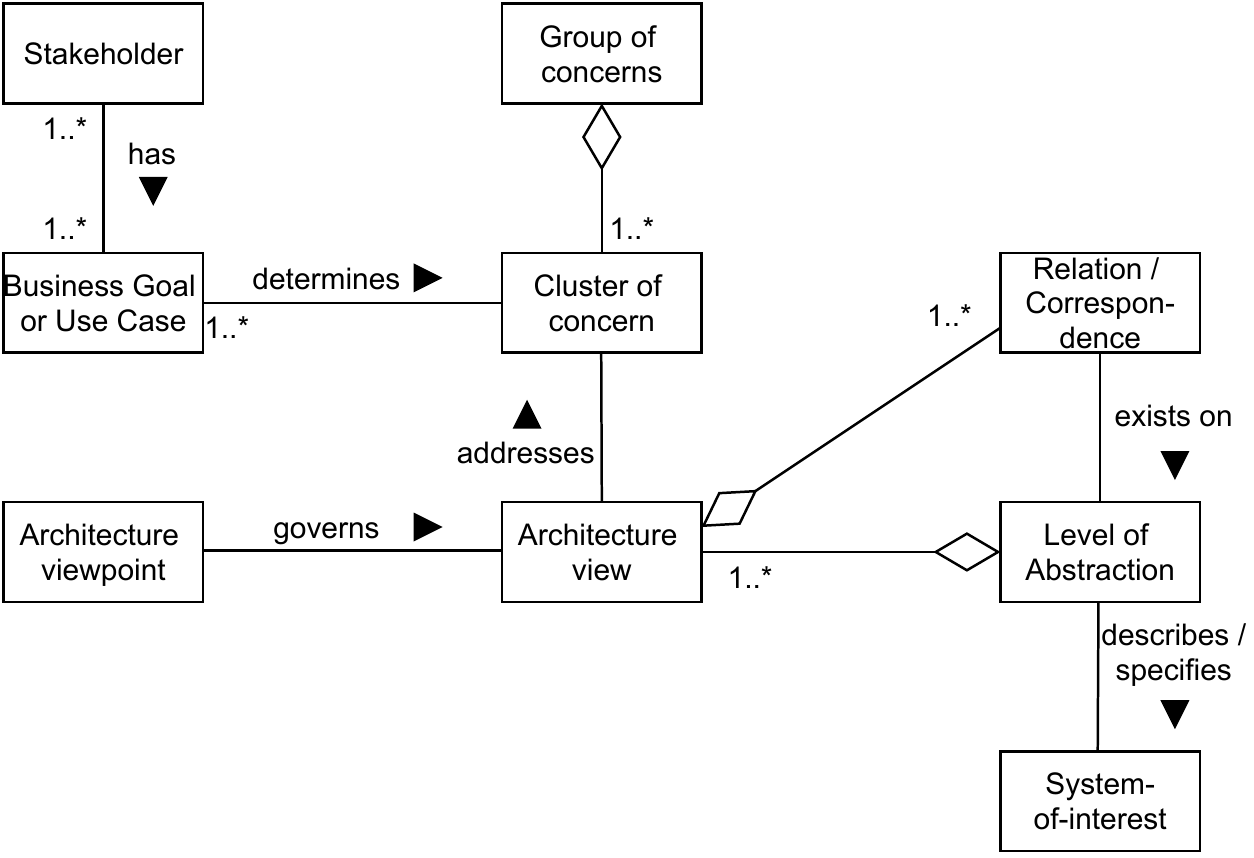} 
\caption{Conceptual model of the compositional architecture framework for VEDLIoT~\cite{Heyn2023}}
\label{fig:architecture_uml}
\vspace{-2.0em}
\end{figure}

The requirement concept is built upon an architecture framework, providing a reusable knowledge structure for VEDLIoT system design. Architecture frameworks organize architectural descriptions and associated requirements into distinct architectural views~\cite{Pelliccione2017}. These different views are necessary to describe the diverse use cases and concerns associated with the VEDLIoT platform. An architectural view expresses ``the architecture of a system from the perspective of specific system concern''~\cite{ISO42010} and can be viewed at different abstraction levels. A hierarchical design process allows the co-evolution of requirements and architecture~\cite{Cleland2013}.
We employ compositional thinking by establishing abstraction levels for architectural views, corresponding requirements, and their classification into clusters of concern. Architectural views within each cluster can be sorted by their represented abstraction level. A conceptual model of the compositional architecture framework approach for VEDLIoT is depicted in \autoref{fig:architecture_uml}. Four abstraction levels are used for VEDLIoT use cases: Knowledge and Analytical level, Conceptual level, Design level, and Run time level.
Required clusters of concern are determined through identified use cases based on the operational context, and high-level goals for the AI system. For example, privacy might not be of concern for an AI-based diagnostic system detecting faults of a welding robot, but safety could be of paramount concern. Relevant clusters of concerns for quality aspects of an AI system in the IoT are Safety, Security, Privacy and Ethical Aspects such as Fairness and Transparency. Additionally, Energy Efficiency can be considered as an explicit quality aspect for embedded systems. Compositional thinking through the proposed architecture framework enables co-designing the system to fulfil identified quality concerns. Early in the development process, correspondences between views regarding quality concerns and other views in the architecture description are established, resulting in a system that is safe, secure, efficient, or fair by design. A comprehensive list of relevant clusters of concerns and architecture views for VEDLIoT can be found in ~\cite{Heyn2023}.

\subsection{Safety Aspects}
\label{sec:safe_sec:safe}
The VEDLIoT approach to Requirements Engineering facilitates constructive design of safety critical systems.
In the architectural framework, a specific cluster of concern is reserved to safety argumentation, enabling safety arguments to be directly connected to architectural views, such as linking hardware design decisions with data quality concerns to support desired safety levels.

Covering both design and run-time aspects of distributed ML in IoT when discussing safety is essential but practically challenging \cite{Heyn2023a}. Challenges like \emph{unsuitable safety standards} and\emph{ missing guidelines for data selection} relate to both challenges of specifying training data and runtime monitoring. These challenges are interdependent and warrant a combined investigation. For instance, \emph{IP protection} causes the \emph{inability to access inner states} and \emph{creating failure models for ML models}.
We recommend the following good practices for safety argumentation in system development involving distributed DL:
\emph{(1)} avoid restrictive IP protection; \emph{(2)} relate confidence measures to actual performance metrics; \emph{(3)} overcome grown data selection habits; and \emph{(4)} balance hardware limitations in embedded systems.
The architectural framework introduced for VEDLIoT-based systems facilitates the implementation of these recommendations.

\subsection{Trusted execution for AIoT}
\label{sec:safe_sec:tee}

Trusted computing in AIoT presents challenges due to the complexities of distributed architectures.
However, secure execution environments for critical software components and intellectual property protection are necessary.
VEDLIoT establishes a cross-platform secure haven using Trusted Execution Environments (TEEs) and focuses on securing server and small device infrastructures across the cloud-edge continuum.
Two secure runtimes have been developed for Intel and ARM, providing comprehensive protection for AIoT systems.

Intel's Software Guard eXtensions (SGX) and ARM's TrustZone create protected memory regions, segregating critical software components from the system via hardware-based isolation.
Despite their advantages, developing and deploying trusted components for these technologies is challenging due to constraints and divergent software paradigms.
VEDLIoT introduces two trusted runtimes, \textsc{Twine}~\cite{DBLP:conf/icde/MenetreyPFS21} for SGX and \textsc{WaTZ}~\cite{menetrey2022watz} for TrustZone, which simplify software creation for secure environments by abstracting TEE complexities.
Both runtimes use WebAssembly, a versatile bytecode format suited for constrained environments, and its modular system interface (WASI), streamlining development and fostering efficient, unified solutions for trusted computing in cloud and IoT environments~\cite{10.1145/3526059.3533618}.

Integrating deep learning frameworks, such as Tensorflow Lite, into secure environments is crucial for bridging the gap between TEEs and AIoT.
WebAssembly enables this integration, facilitating secure AIoT application development.
In a smart mirror use-case, voice data is processed within a TEE, ensuring robust protection against eavesdropping and malicious software interference while preserving the deep learning model's confidentiality.
This highlights the potential to enhance security and privacy in AIoT applications through the integration of TEEs and WebAssembly technology.

\subsection{Inferring trust in AIoT}

While TEEs address confidentiality and integrity concerns, verifying the authenticity of executing software remains crucial.
Remote attestation (RA) ensures the trustworthiness of software, hardware, and data on remote devices and facilitates secure communications among distributed components~\cite{menetrey2022dais}.
VEDLIoT has contributed to several RA-based solutions.

Integrating RA capabilities into trusted WebAssembly runtimes enhances their security.
For Intel, the system uses SGX features for enclave generation and attestation.
But TrustZone lacks inherent attestation capabilities.
We addressed this by leveraging protected keys within the SoC die, exclusively accessible by the TEE, and developing a remote attestation protocol for WebAssembly applications based on the Sigma protocol.
This enables both platforms to provide attested proof of application trustworthiness and facilitate secure communication and confidential data exchange for remote parties.

To bolster remote attestation robustness, VEDLIoT decentralizes attestation logic, enhancing resilience to Byzantine attacks.
We developed SIRE, a Byzantine fault-tolerant infrastructure supporting remote attestation, extensible coordination, application membership management, and auditable integrity-protected logging.
SIRE combines multiple functionalities to address IoT challenges like heterogeneity, untrustworthiness, and high dynamism.
Combining remote attestation with Byzantine fault tolerance strengthens the attestation process, providing more robust security guarantees.
Furthermore, SIRE's additional functionalities, such as coordination primitives, facilitate complex task execution and simplify application management and deployment.

 \section{VEDLIoT Applications}
\label{sec:apps}

\subsection{Automotive}
The Automotive use case aims to optimize processing speed, energy efficiency and minimize on-car energy consumption. Pedestrian Automatic Emergency Breaking (P-AEB) is a well-defined safety application in the Automotive space. Based on sensor information a decision is taken to break the vehicle if a pedestrian is on or close to the travel path of the vehicle. AI/DL has been used to solve the end-to-end processing over the full detection-decision-actuation space. Multiple different AI models have been used for state-of-the-art product implementations. In the current Veoneer use-case evaluation we have implemented a version of EfficientNet.

The main challenge of this automotive use-case is the limited processing power available in the vehicle, the communication resource limits and the possible processing power available in the edge, in this case a base station. In the VEDLIoT project we have evaluated ways of distributing the DL model over multiple processing nodes, some with minimal and some with high capabilities,  to reduce the system response time. We have optimized the partitioning of the DL model, based on the total end-to-end latency, the available cellular communication capacity and the edge processing availability and speed. As the vehicle is constantly moving, the communication and edge processing resources will vary partly due to the number of other users of the cellular system and partly due to the distance to the base-station which potentially may lower the communication throughput.

We have used the Architecture and Requirement Frameworks to identify the necessary system components. Based on this we have designed a prototype/demo system based on a u.RECS platform to mimic the sensor (in this case an RGB camera) processing part plus the vehicle central compute unit (CCU). The u.RECS is hosting both a SMARC device (= sensor processing unit) and an NVIDIA Jetson (= CCU). The CCU is connected, through a standard 4g connection, to a base station which will be equipped with a tRECS hardware unit.

Veoneer has done data collection with a camera system based on the performance requirements identified with the Requirements Framework. This data set was then used for the DL model learning and testing phases. The DL model optimization tool, developed within VEDLIoT, was then used to distribute the DL model over all of the processing nodes described above. Multiple optimization scenarios, depending on the dynamically changing vehicle environment, was designed and will be used to adapt and re-configure the processing system as the vehicle drives through the traffic scene.

Since both data and DL models are sent over open communication links, security, robustness, trust and safety have been investigated within the VEDLIoT project focusing on the extremely high functional safety requirements set by critical automotive applications. Comparisons of different hardware setups using either only the sensor processing unit, the sensor processing unit and the central compute unit and finally both vehicle processing units and an edge unit, connected by a wireless link, have been performed showing that the latency may be lowered using all available processing devices while still maintaining the same AI/DL processing accuracy. The next step is to do comparisons of the total energy consumption for the different setups, including the communication power consumption.
\subsection{Industrial IoT}
Industrial use cases can also benefit from the integration of DL algorithms. The two use cases supported by VEDLIoT, Motor Condition Classification and Series Arc Fault Detection (AFD) in low-voltage direct current (LVDC) systems, represent two categories of industrial problems: predictive maintenance and anomaly detection. Challenges of industrial use cases and especially AI-based solutions are of various aspects: limited data of fault condition under real scenarios due to finance and safety considerations, and the lack of framework for software development with DL. VEDLIoT addresses these challenges and accelerates the development of both industrial use cases. Both use cases are first investigated with the requirement and security analysis framework from VEDLIoT. This determines the potential risks in implementing AI-based solutions and the specification for data and model, which guide the system design for better performance.

Based on the analysis and system design, test-benches are built and gradually improved for condition simulation and data generation. AI accelerators are integrated in the systems, and the first models are trained and validated on the test-benches. The test-benches provide not only the environment for validation and demonstration, including hardware efficiency, model accuracy, and effectiveness of analysis framework, but also a playground for AIoT technologies and protocols such as secure communication and integration of augmented reality (AR).

The focus of Motor Condition Classification use case is energy efficiency. It aims to build an easy-to-mount, battery-powered DL-driven sensor box for on-site data processing, to monitor thermal, operational, and mechanical conditions of motors. 
With the insight in accelerator architecture from VEDLIoT \cite{vedliot_d33}, the AI accelerator MAX78000 is selected and integrated in the sensor box with temperature, vibration, and magnetic flux sensors on our customized hardware. The sensor box is capable of data collection, processing with DL model, and transmission within secure network environment. The data is then visualized in HMI (human machine interface) with AR technology. The goal of the other use case AFD in LVDC systems is to develop DL-based solution for fast arc detection with high accuracy. Based on the analysis, this use case is more time and accuracy critical compared to the motor use case. Therefore, a more powerful AI accelerator NVIDIA Jetson Xavier NX is implemented for the prototype. For the arc detection, current signal from the circuit in the arc test-bench is collected by an ADC (analogue digital converter) with 16kHz sampling rate and transmitted to the accelerator for model inference. The time interval between the data generation and the delivery of classification result is on average 10ms. The model can also reach over 95\% accuracy based on the collected data.

\subsection{Smart home}
In VEDLIoT, as part of the smart home use case, a smart mirror was developed  with a focus on data protection through local processing. The mirror provides an intuitive user interface and displays personalized information as well as the status of the smart home. While user identification is performed by facial recognition and tracking using depth imaging cameras, hand gestures are recognized and used to control the mirror. Furthermore, a voice assistant, supported by natural language processing (NLP) allows the user to control key features through voice commands. All computations of the smart mirror are executed locally on the device using open-source software, ensuring maximum privacy.
\begin{figure}[!h]
\centering
\includegraphics[width=0.93\columnwidth]{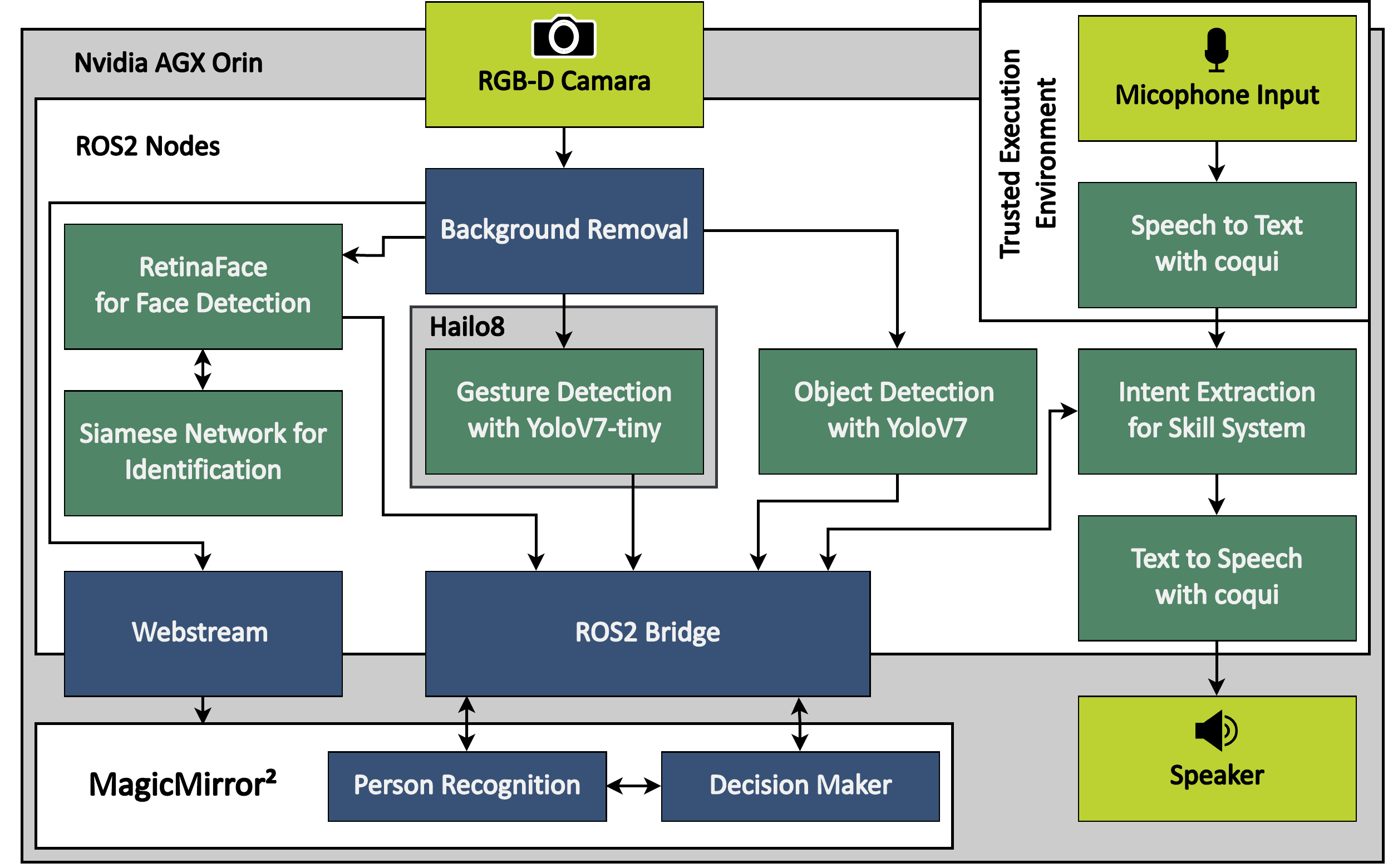} 
\caption{Data flow within the smart mirror demonstrator}
\label{fig:magic_mirror_pipeline}
\vspace{-1.0em}
\end{figure}
To achieve the necessary detections, the project combines multiple machine learning techniques. Currently, the software architecture is deployed on one NVIDIA Orin AGX, combined with an M.2-based Hailo-8 DL accelerator for gesture detection. The addition of the Hailo-8 ASIC accelerator significantly improves the system performance, such that it ensures stable 30~FPS and decreases overall power consumption to only 49~Watts. The local voice assistant is implemented on the far edge computing platform u.RECS. Spoken keywords are recognized by a hot word detection, which triggers an encrypted audio-stream towards the local edge-server, ensuring the highest level of user privacy. The next steps in the development will be the deployment of all smart mirror components onto the newly developed u.RECS platform. The goal is to execute the language and vision processing pipelines on a single system, e.g., through the combination of an NVIDIA Orin NX and a Hailo-8 accelerator.

 \section{Summary}
\label{sec:summary}
VEDLIoT tackles the issue of integrating Deep Learning into IoT devices with restricted computing capabilities and minimal power consumption requirements, raising the need for energy-efficient computing. The VEDLIoT AIoT hardware platform offers tailored hardware components and supplementary accelerators for AIoT applications, ranging from embedded systems to edge computing and cloud platforms. An efficient middleware simplifies neural network programming, testing, and deployment to this diverse hardware ecosystem. Innovative approaches for requirements engineering, combined with safety and security principles, address the challenges posed by implementing Deep Learning techniques throughout the entire framework. These concepts are validated through rigorous use cases in vital industry sectors such as automotive, automation, and smart home.
 
\section*{Acknowledgments}
This publication incorporates results from the VEDLIoT project, which received funding from the European Union's Horizon 2020 research and innovation programme under grant agreement number 957197.

\printbibliography

\end{document}